\documentclass[twocolumn]{aastex62}

\newcommand{\planck}{\textsl{Planck}}

\newcommand{\lcdm}{\ensuremath{\Lambda\mathrm{CDM}}}
\newcommand{\wmap}{\textsl{WMAP}}

\usepackage{bm} 
\usepackage{mathtools}

\graphicspath{{./}{figures/}}

\shorttitle{Deconstructing TT Power Spectrum}
\shortauthors{Kable et al.}

\begin{document}

\title{Deconstructing the \planck\ TT Power Spectrum to Constrain Deviations from \lcdm}

\correspondingauthor{Joshua A. Kable}
\email{jkable2@jhu.edu}

\author[0000-0002-0786-7307]{Joshua A. Kable}
\affil{Johns Hopkins University \\
3400 North Charles Street \\
Baltimore, MD 21218, USA}

\author{Graeme E. Addison}
\affiliation{Johns Hopkins University \\
3400 North Charles Street \\
Baltimore, MD 21218, USA}

\author{Charles L. Bennett}
\affiliation{Johns Hopkins University \\
3400 North Charles Street \\
Baltimore, MD 21218, USA}

\begin{abstract}
Consistency checks of \lcdm\ predictions with current cosmological data sets may illuminate the types of changes needed to resolve cosmological tensions. To this end, we modify the CLASS Boltzmann code to create phenomenological amplitudes, similar to the lensing amplitude parameter $A_L$, for the Sachs-Wolfe, Doppler, early Integrated Sachs-Wolfe (eISW), and Polarization contributions to the CMB temperature anisotropy, and then we include these additional amplitudes in fits to the \planck\ TT power spectrum. We find that allowing one of these amplitudes to vary at a time results in little improvement over \lcdm\ alone suggesting that each of these physical effects are being correctly accounted for given the current level of precision. Further, we find that the only pair of phenomenological amplitudes that results in a significant improvement to the fit to \planck\ temperature data results from varying the amplitudes of the Sachs-Wolfe and Doppler effects simultaneously. However, we show that this model is really just refinding the \lcdm\ + $A_L$ solution. We test adding our phenomenological amplitudes as well as $N_{\textrm{eff}}$, $Y_{\textrm{He}}$, and $n_{\textrm{run}}$ to \lcdm\ + $A_L$ and find that none of these model extensions provide significant improvement over \lcdm\ + $A_L$ when fitting \planck\ temperature data. Finally, we quantify the contributions of both the eISW effect and lensing on the constraint of the physical matter density from \planck\ temperature data  by allowing the phenomenological amplitude from each effect to vary. We find that these effects play a relatively small role (the uncertainty increases by $3.5\%$ and $16\%$ respectively) suggesting that the overall photon envelope has the greatest constraining power. 
\end{abstract}


\section{Introduction}
\lcdm\ is the standard model of cosmology because with only six parameters, it successfully explains a wide range of cosmological and astrophysical phenomena. However, in recent years, tensions have emerged between the preferred values of cosmological parameters resulting from fits to cosmological data sets assuming the \lcdm\ model and direct measurements of those cosmological parameters. In particular, there is a 4.4$\sigma$ tension in the preferred value of the Hubble constant, $H_0$, between the cosmological distance ladder measurement by SH0ES, $H_0 = 74.02 \pm 1.42$ km$\textrm{s}^{-1}\textrm{Mpc}^{-1}$ \citep{Riess/etal:2019}, and the inferred value from the most precise measurements to date of the Cosmic Microwave Background (CMB) provided by \planck, $H_0 = 67.37 \pm 0.54$ km$\textrm{s}^{-1}\textrm{Mpc}^{-1}$ \citep{planck/6:2018}. 

The $H_0$ tension can be divided into a discordance between the preferred values by early universe observations assuming \lcdm\ and direct measurements in the late universe. While this tension is usually expressed as a disagreement between \planck\ and the cosmological distance ladder, \cite{addison/etal:2018} show that a similar discordance is found when combining Baryon Acoustic Oscillation (BAO) data with \planck\ CMB measurements, CMB measurements from experiments other than \planck, or with primordial deuterium abundances using no CMB anisotropy data \citep[see also, e.g.,][]{aubourg/etal:2015,Cuceu/etal:2019,eBoss/2020}. 

On the late universe side, this tension persists even if different calibrators are used for the cosmological distance ladder \citep{Huang.etal:2020}. Using the Tip of the Red Giant Branch as a calibrator results in $H_0 = 69.6 \pm 1.9$ km$\textrm{s}^{-1}\textrm{Mpc}^{-1}$ \citep{freedman/etal:2019}, but \cite{Yuan/etal:2019} argue that this analysis overestimates the Large Magenlanic Cloud extinction and instead determine $H_0 = 72.4 \pm 2.0$ km$\textrm{s}^{-1}\textrm{Mpc}^{-1}$. Completely independent of the cosmological distance ladder, strong gravitational lensing time delays by $H_0$ Lenses in COSMOGRAIL's Wellspring (H0LiCOW) determine $H_0 = 73.3 \pm 1.7$ km$\textrm{s}^{-1}\textrm{Mpc}^{-1}$, which is in 3.9$\sigma$ tension with \planck\ \citep{Wong/etal:2020}. 

Because the $H_0$ tension exists between multiple data sets and breaks down by cosmological epoch instead of observational technique, it is unlikely to be resolved by an underestimated or unmodeled systematic, suggesting the need for physics beyond the standard model of cosmology. Finding extensions to \lcdm\ that resolve the Hubble tension yet stay consistent with the multitude of cosmological data sets is challenging \citep[see, e.g.,][]{Knox/etal:2020}. For example, it has been proposed that incorporating a form of dark energy that comprises about 10$\%$ of the energy density of the universe around matter-radiation equality and then decays away before recombination can alleviate the $H_0$ tension \citep{Poulin/etal:2019,Lin/etal:2019,Berghuas/etal:2020}. However, fitting these current Early Dark Energy models to \planck\ data results in an increase in the cold dark matter density that is disfavored by large scale structure measurements \citep{Hill/etal:2020,D'Amico/etal:2020,Ivanov/etal:2020}. 

In the absence of a clear theoretical direction, it can be useful to perform consistency checks of \lcdm\ predictions with current data sets to determine what kinds of changes to the standard model are necessary or even allowed \citep[see, e.g.][]{Kable/etal:2020,Motloch:2020}. It has been shown that in addition to the $H_0$ tension with direct measurements, \planck\ data prefers a 2-3$\sigma$ larger value of $S_8 = \sigma_8\sqrt{\Omega_m/0.3}$, which measures matter clustering, than weak lensing experiments \citep{Hildebrandt/etal:2018,Joudai/etal:2018,Abbott/etal:2018,Hikage/etal:2019} and clustering abundance surveys \citep[e.g.,][]{Lin/etal:2017,McCarthy/etal:2018}. 

Additionally, there is a $\sim$2.5$\sigma$ tension between the preferred values of parameters like the physical cold dark matter density, $\omega_c$, for \planck\ TT $\ell \leq 1000$ and \planck\ TT $\ell > 1000$ (or similarly for \planck\ TT $\ell \leq 800$ and \planck\ TT $\ell > 800$), which can be resolved by allowing the amplitude of the lensing contribution to the CMB TT power spectra to vary \citep[e.g.,][]{addison/etal:2016,planck/6:2018}. This is done by extending \lcdm\ to include a phenomenological amplitude, $A_L$, which rescales the amplitude of the lensing power spectrum as
\begin{align}
    C_{\ell}^\Psi\rightarrow A_LC_{\ell}^\Psi,
\end{align}
where $A_L$ has a physical value of 1 \citep{calabrese/etal:2008}. The combined \planck\ TT, TE, and EE power spectra prefer $A_L > 1$ at 2.8$\sigma$, which is driven largely by an improvement to the fit for multipoles $1100 \leq \ell \leq 2000$ in the \planck\ TT power spectrum, though there is also improvement to the fit for \planck\ TT $\ell < 30$ \citep{planck/6:2018}. However, the lensing power spectrum reconstructed from higher-order statistics of the \planck\ maps is in good agreements with standard \lcdm\ predictions \citep[e.g.][]{Simrad/etal:2018,Motloch/etal:2020,planck/VIII:2018}. Moreover, the South Pole Telescope Polarimeter (SPTpol) TE and EE power spectra prefer $A_L < 1$ at 1.4$\sigma$, and the Atacama Cosmology Telescope (ACT) DR4 is consistent with $A_L = 1$ within $1\sigma$ \citep{Henning/etal:2018,ACTDR4}. While the \planck\ TT power spectrum prefers greater peak smoothing consistent with $A_L > 1$, other cosmological data sets disfavor changing the physical amount of lensing. 

In this paper, we create phenomenological amplitudes analogous to $A_L$ for the early Integrated Sachs-Wolfe (eISW), Sachs-Wolfe, Doppler and the Polarization effects \footnote{Note that this polarization effect refers to the the contribution to the total intensity that is sourced by CMB polarization. We define this in more detail in Section 2.}, which all source the CMB temperature anisotropy. We fit these new phenomenological amplitudes to \planck\ data to determine if there are any deviations from standard \lcdm\ favored by \planck. In this way, we deconstruct the TT power spectrum into its constituent sources, which provides a test for where potential model extensions are allowed or are necessary. While scaling these physical effects can affect the CMB TE power spectrum, we choose to fit only the \planck\ TT power spectrum as we are primarily interested in quantifying deviations from \lcdm\ predictions in the temperature anisotropy, which is already known to have internal differences in the preferred \lcdm\ parameter values between \planck\ TT $\ell \leq 800$ and \planck\ TT $\ell > 800$.  The \planck\ Collaboration performed a similar exercise and found that these phenomenological amplitudes are consistent with expectations \citep[see footnote 30 of][]{planck/6:2018}. We quantify this consistency and extend the analysis to include combinations of the phenomenological amplitudes.

There is a well-known degeneracy in the CMB temperature data between the scalar amplitude, $A_s$, and the optical depth, $\tau$. This degeneracy is broken by the reionization bump measured by $\ell \lesssim 20$ EE data \citep[see e.g. Figure 8][]{planck/V:2019}. For all cases in this paper, we include a Gaussian prior of $\tau = 0.0506 \pm 0.0086$ to account for the constraint from Planck Low $\ell$ EE data as described by \cite{planck/6:2018}. We tested the impact of changing both the mean value and width of the Gaussian prior on $\tau$ and found that our conclusions were insensitive to these changes. 

This paper is organized as follows. In Section 2 we define the phenomenological amplitudes for the eISW, SW, Doppler, and Polarization effects and discuss how each phenomenological amplitude affects the TT power spectrum. In Section 3 we show the constraints provided by the \planck\ 2018 TT power spectrum when we allow one or more of the phenomenological amplitudes to vary. In Section 4 we test possible extensions to \lcdm\ + $A_L$ to determine where if any further improvement in the fit can be found. Finally in Section 5, we provide conclusions. 

\section{The Phenomenological Amplitudes}
\subsection{Definitions of Phenomenological Amplitudes}
In this section, we define the phenomenological amplitudes that we will use for the rest of the paper. The perturbation away from a blackbody spectrum of the CMB photon distribution, $\Theta \equiv \delta T/T$, can be quantified by integrating the various cosmological perturbations along the path of the photons. This distribution can be expanded in terms of spherical Bessel functions, $j_{\ell}$, and wavenumbers, k, for a given perturbation as
\begin{equation} \label{eq2}
\begin{split}
    &\Theta_{\ell}(k,\eta_{0}) = A_{\textrm{SW}}\int_0^{\eta_0}d\eta g(\eta) \Big[\Theta_0(k,\eta) + \Psi(k,\eta)\Big]j_{\ell}\big[k(\eta-\eta_0)\big]\\
    &- A_{\textrm{Dop}}\int_0^{\eta_0}d\eta g(\eta)\frac{iv_b}{k}\frac{d}{d\eta}j_{\ell}\big[k(\eta-\eta_0)\big]\\
    &+ \int_0^{\eta_0}d\eta f(z(\eta),A_{\textrm{eISW}})e^{-\tau}\Big[\dot{\Psi}(k,\eta) - \dot{\Phi}(k,\eta)\Big]j_{\ell}\big[k(\eta-\eta_0)\big]\\
    &+ A_{\textrm{Pol}}\int_0^{\eta_0}d\eta \Big[\frac{g(\eta)\Pi}{4} + \frac{3}{4k^2}\frac{d^2}{d\eta^2}\big[g(\eta)\Pi\big]\Big]j_{\ell}\big[k(\eta-\eta_0)\big],
\end{split}
\end{equation}
following \cite{ModernCosmology}. In this equation, $\eta$ is conformal time, $\tau$ is the optical depth at a given conformal time, $g(\eta)\equiv -\dot{\tau}e^{-\tau}$ is the visibility function, $\Psi$ is the Newtonian potential, $\Phi$ is the spatial perturbation to the metric, $v_b$ is the velocity of the baryons, and $\Pi$ is the polarization tensor. 

The visibility function is a probability density of the conformal time when a CMB photon last scattered, so it is sharply peaked around recombination. This in turn means that the first, second, and fourth terms are sourced primarily at the surface of last scattering while the third term is sourced at all points along the way. 

The first term accounts for the Sachs-Wolfe effect, which is the relative redshifting or blueshifting of CMB photons as they leave the last scattering surface due to fluctuations in the size of the  gravitational potential wells. The second term accounts for the Doppler shifting of CMB photons moving toward or away from the observer along the line of sight. The third term is the contribution of the ISW effect. Much like the Sachs-Wolfe effect, this quantifies the redshifting and blueshifting of CMB photons as they climb out of and fall into gravitational potential wells; however, in this case the size of the potential wells decays because of either radiation in the early universe or dark energy in the late universe. The final term is the CMB polarization contribution to the CMB temperature anisotropy. This results from the directional dependence of Compton scattering and the coupling of the CMB polarization to the quadrupole moment of $\Theta$, which is discussed by \cite{Hu/etal:1996}.

In Equation 2, we have defined phenomenological amplitudes for each of these effects. We adopt the same convention as \cite{Hou/etal:2013} where the phenomenological amplitudes scale the sources of the photon distribution. Additionally, we define $f(z(\eta),A_{\textrm{eISW}}) = A_{\textrm{eISW}}$ when z $>$ 30 and unity for z $\leq$ 30 as was done in \cite{Hou/etal:2013}. We could additionally define a phenomenological amplitude to account for the late time ISW effect, but we find that this is too poorly constrained by the CMB data to provide a meaningful test. 

\subsection{Effects of Varying Phenomenological Amplitudes on Theory TT Power Spectrum}
Before we discuss results of extending \lcdm\ to include these phenomenological amplitudes when fitting to \planck\ TT data, we illustrate the general effects on the TT power spectrum of varying each of these phenomenological amplitudes. To do so, we modify the source function in the CLASS Boltzmann code \citep{CLASS/1:2011,CLASS/2:2011} to include these new parameters. In Figure 1, we show the effect on the CMB power spectrum of changing each of the four phenomenological amplitudes as well as $A_L$. In each case, we employ a fiducial cosmology resulting from a Markov Chain Monte Carlo (MCMC) using Monte Python \citep{Audren/etal:2013,Brinckmann/etal:2018} of \planck\ 2018 TT data with a prior of $\tau = 0.0506 \pm 0.0086$.

\begin{figure*}[!tbp]
  \begin{minipage}[b]{0.4\textwidth}
    \hspace*{-1.25cm}
    \includegraphics[width=8.5in, height=7.5in]{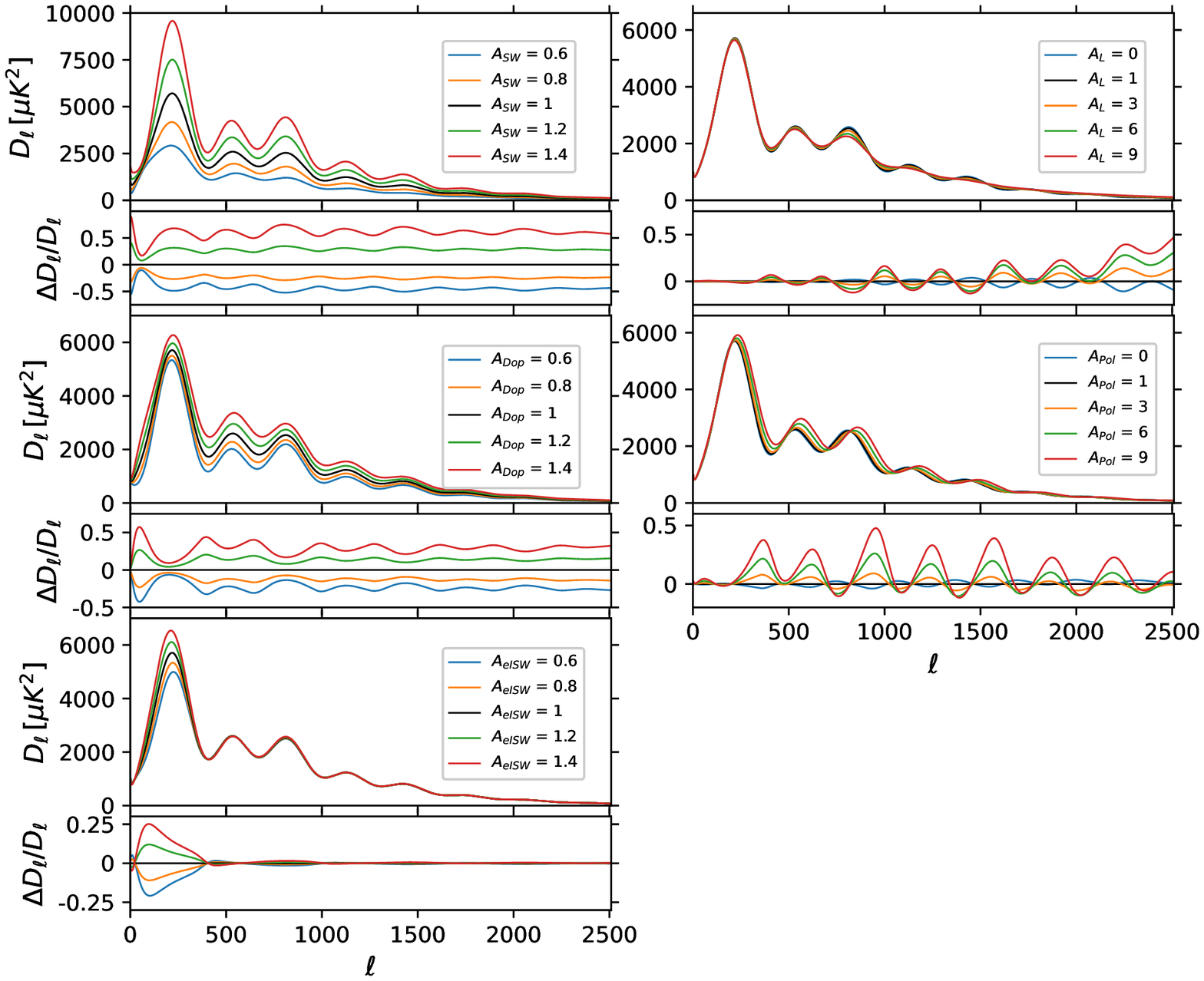}
  \end{minipage}
  \caption{Impact of varying a range of phenomenological amplitudes, including $A_L$ and the phenomenological amplitudes defined in Equation 2, while fixing the \lcdm\ parameters. We additionally show the ratio of the residual with the \lcdm\ case below each plot. In this paper, we examine the consistency of these amplitudes with unity as a consistency check of the standard model. If the data show a significant preference for the phenomenological amplitude not equal to unity, then this gives a clue as to what physics an alternative cosmological model would have to alter to match the data better than \lcdm.  
 \label{fig:Fig2}}
  \end{figure*}
  
Changing $A_{\textrm{SW}}$ has the largest effect on the overall amplitude of the power spectrum of the parameters varied in Figure 1. While increasing $A_{\textrm{SW}}$ increases both acoustic peaks and troughs, it increases the heights of the peaks by a larger fraction. The effect is stronger on the compression modes (odd peaks), where the baryon-photon fluid is at the bottom of the gravitational potential, than the rarefaction modes (even peaks). Increasing the Sachs-Wolfe effect leads to deeper potentials allowing for greater compression. Increasing $A_{\textrm{SW}}$ also leads to a small phase shift toward larger scales.
 
Increasing $A_{\textrm{Dop}}$ also results in an overall increase to the power spectrum and a small phase shift to larger scales, but unlike the Sachs-Wolfe effect, it disproportionately  impacts the troughs and even peaks of the power spectrum. In particular, the ratio of the heights of the peaks to troughs decreases as $A_{\textrm{Dop}}$ increases. The Doppler effect is proportional to the baryon velocity as shown in Equation 2. In the absence of baryon loading, the baryon velocity would peak when $\Theta_o + \Psi = 0$, which corresponds to the troughs of the CMB power spectrum \citep[see, e.g., Section 5.2 of ][]{Hu/Thesis}. With the baryon loading, the baryon velocity still peaks near the troughs and therefore increasing $A_{\textrm{Dop}}$ fills in the troughs. The rarefaction modes get more power than the  compression modes because increasing the baryon velocity increases the pressure, which makes it easier for photons to escape the gravity wells. 

Changing $A_{\textrm{eISW}}$ primarily affects the first peak, but also makes small contributions to the higher peaks with a preference for the odd acoustic peaks. $A_{\textrm{eISW}}$ has the largest effect on the first acoustic peak because it has the largest effect on modes that enter the horizon when the universe is dominated by matter but still has a sizable radiation density \citep[see, e.g., Section 8.6 of ][]{ModernCosmology}. Increasing $A_{\textrm{eISW}}$ causes an increase in power because it increases the radiation density, which hastens the decay of the gravity wells. There is also a slight filling in of the second trough. 

Finally, Figure 1 shows that changing $A_{\textrm{Pol}}$ makes the smallest change to the amplitude of the power spectrum. Increasing $A_{\textrm{Pol}}$ results in a phase shift to smaller scales. This phenomenological amplitude is coupled to the CMB quadrupole moment, $\Theta_2$, which sources photon diffusion damping \citep[see, e.g., Section 8.4 of][]{ModernCosmology}. Hence, increasing $A_{\textrm{Pol}}$ results in increased damping.

\section{Results from varying phenomenological amplitudes}
In the previous section, we defined phenomenological amplitudes for the Sachs-Wolfe, eISW, Doppler, and Polarization effects that source the CMB temperature anisotropy. In this section, we explore how these phenomenological amplitudes are constrained by the CMB by running MCMC fits on \planck\ 2018 TT data. To sample the posterior distributions for the model parameters, we use our modified CLASS Boltzmann code, which includes the amplitudes defined in Equation 2 as additional model parameters, and run MCMCs using Monte Python. 

We use the likelihoods for \planck\ 2018 TT High $\ell$ Lite corresponding to $30 \leq \ell \leq 2508$ and \planck\ 2018 TT Low $\ell$ corresponding to $\ell < 30$ provided by the \planck\ Collaboration. We choose to use the Lite likelihoods, where foreground parameters have already been marginalized over, because we are not investigating the impact of altering the foreground model in this work. Hereafter, we will refer to this likelihood as \planck\ TT.

For certain models, we also explore splitting the \planck\ data to highlight the discrepancy between the parameter posteriors resulting from sampling \planck\ TT $\ell \leq 800$ and \planck\ TT $\ell > 800$. We choose to split the \planck\ data at $\ell = 800$ because this corresponds to the point where each split of the \planck\ data has roughly equivalent constraining power \citep[e.g.][]{planck/51:2017}. We refer to these data split likelihoods as \planck\ TT $\ell \leq 800$ and \planck\ TT $\ell > 800$ respectively. 

Finally, unless otherwise specified, we use a Gelman-Rubin convergence statistic of $R-1 = 0.05$ for the least constrained parameter to define the point when our MCMC chains have converged \citep{gelman/rubin:1992}. 

\subsection{Fits to \lcdm\ Plus One Phenomenological Amplitude}
In this subsection, we compare the MCMC fits to \planck\ TT assuming \lcdm\ + one phenomenological amplitude to the MCMC fits to \planck\ TT assuming \lcdm. The results of these MCMC fits are summarized in Table 1 and Figures 2 and 3. In Table 1, we show that no variations of the phenomenological amplitudes that we introduced in Section 2 are able to fit \planck\ TT significantly better than \lcdm. Moreover, no variations of these phenomenological amplitudes are able to alleviate the $H_0$ tension.  

\begin{deluxetable*}{ccccccc}
 \centering
\tablewidth{6.5in}

 \tablecaption{Mean values and 68$\%$ credible intervals for standard \lcdm\ and for \lcdm\ plus one phenomenological amplitude variation for the MCMC chains fit to \planck\ TT. For definitions of the phenomenological amplitudes see Section 2. We use a prior of $\tau = 0.0506 \pm 0.0086$. \label{tab:best-fit}}

\tablehead{
 \colhead{Parameter} & \colhead{\lcdm} & \colhead{$+A_L$} & \colhead{$+A_{\textrm{SW}}$} & \colhead{$+A_{\textrm{Dop}}$} & \colhead{$+A_{\textrm{eISW}}$} & \colhead{$+A_{\textrm{Pol}}$} 
 }

 \startdata
 $H_0$ & 67.00 $\pm$ 0.93 & 69.11 $\pm$ 1.20 & 67.41 $\pm$ 1.03 & 66.90 $\pm$ 1.02 & 66.55 $\pm$ 0.96 &  67.41 $\pm$ 1.27 \\
 $100*\omega_b$ & 2.213 $\pm$ 0.022 & 2.265 $\pm$ 0.029 & 2.226 $\pm$ 0.027 & 2.210 $\pm$ 0.030 & 2.170 $\pm$ 0.035 & 2.225 $\pm$ 0.033\\
 $\omega_c$ & 0.1205 $\pm$ 0.0021 & 0.1164 $\pm$ 0.0025 & 0.1198 $\pm$ 0.0022 & 0.1206 $\pm$ 0.0022 & 0.1206 $\pm$ 0.0021& 0.1206 $\pm$ 0.0021 \\
 $10^9A_se^{-2\tau}$ & 1.8847 $\pm$ 0.0140 & 1.8658 $\pm$ 0.0156 & 1.904 $\pm$ 0.027 & 1.886 $\pm$ 0.022 & 1.873 $\pm$ 0.0165& 1.8836 $\pm$ 0.0147\\
 $n_s$ & 0.9634 $\pm$ 0.0057 & 0.9751 $\pm$ 0.0072 & 0.9666 $\pm$ 0.0067 & 0.9631 $\pm$ 0.0058 & 0.9713 $\pm$ 0.0077& 0.9657 $\pm$ 0.0077 \\
 $A_{new}$ & ---------& 1.259 $\pm$ 0.099 & 0.9909 $\pm$ 0.0100 & 0.9984 $\pm$ 0.0126 & 1.064 $\pm$ 0.042&  1.16 $\pm$ 0.31 \\
 \hline
 $\Omega_mh^2$ & 0.1426 $\pm$ 0.0020 & 0.1390 $\pm$ 0.0023 & 0.1421 $\pm$ 0.0020 & 0.1427 $\pm$ 0.0020 & 0.1423 $\pm$ 0.0020& 0.1428 $\pm$ 0.0020\\
 $\Omega_m$ & 0.3179 $\pm$ 0.0132 & 0.2915 $\pm$ 0.0148 & 0.3129 $\pm$ 0.0141 & 0.3186 $\pm$ 0.0139 & 0.3217 $\pm$ 0.0134& 0.3147 $\pm$ 0.0148 \\
 $\sigma_8$ & 0.8130 $\pm$ 0.0097  & 0.7933 $\pm$ 0.0120 & 0.8147 $\pm$ 0.0099 & 0.8136 $\pm$ 0.0114 & 0.8144 $\pm$ 0.0097& 0.8145 $\pm$ 0.0102 \\
 \hline
 $\chi^2$ & 229.50 & 221.45& 228.59 & 229.49 & 227.24& 229.18 \\
 $\chi^2_{\Lambda CDM} -\chi^2$ & 0 & 8.5& 0.91 & 0.01 & 2.26& 0.32
 \enddata

 \vspace{-0.5cm}

\end{deluxetable*}

Varying $A_{\textrm{eISW}}$ results in the largest improvement over standard \lcdm\ of these new phenomenological amplitudes. Nevertheless, this variation results in a $<$2$\sigma$ shift in the posterior distribution for $A_{\textrm{eISW}}$ away from the fiducial value of 1. Moreover, the difference in $\chi^2$ found by adding $A_{\textrm{eISW}}$ corresponds to a Probability To Exceed (PTE) of 0.13 further indicating that including $A_{\textrm{eISW}}$ is not a significant model improvement over \lcdm. Considering that we tested four model extensions to standard \lcdm, it is not surprising that one of them resulted in a $>1\sigma$ improvement to the fit. 

To understand where this minor improvement is coming from, we fit \lcdm\ + $A_{\textrm{eISW}}$ to \planck\ TT but excluded multipoles $\ell < 30$ and found that the preference for $A_{\textrm{eISW}} > 1$ was reduced to $<0.5 \sigma$. This suggests that the primary improvement over \lcdm\ when fitting \lcdm\ + $A_{\textrm{eISW}}$ to \planck\ TT comes from multipoles $\ell < 30$. In particular, we find that the TT power spectrum resulting from the best-fit cosmology for \lcdm\ + $A_{\textrm{eISW}}$ has less power than standard \lcdm\ for $\ell < 30$ when fit to \planck\ TT, which allows this model to fit the well-known deficit of power at $\ell < 30$ in \wmap\ and \planck\ TT data \citep{bennett/etal:2013,planck/6:2018}. When $A_{\textrm{eISW}}$ is allowed to vary, \planck\ TT prefers a decrease in the preferred value of $A_s$ and an increase in the preferred value of $n_s$, which results in a reduction in power for $\ell < 30$ for the the best-fit TT power spectrum. 

From Table 1, we see that the improvement found by \lcdm\ + $A_{\textrm{eISW}}$ over standard \lcdm\ is primarily compensated by a 0.043 shift downward in the value of $100*\omega_b$ (100 times the physical baryon density), which corresponds to almost twice the original uncertainty. On a related note, the uncertainty of the baryon density when varying the amplitude of the eISW effect increases by roughly 60$\%$, which illuminates how powerful the relative peak heights, and in particular the height of the first acoustic peak, are in constraining the physical baryon density. 

After $A_L$ and $A_{\textrm{eISW}}$, allowing $A_{\textrm{SW}}$ to vary results in the next most significant improvement over just \lcdm, which can be seen by the approximately 1$\sigma$ shift in the value of $A_{\textrm{SW}}$ from the fiducial value of unity. While the uncertainties on the \lcdm\ parameters increase, such as the near doubling of the uncertainty of $A_se^{-2\tau}$, most parameter shifts are $<0.5 \sigma$. Adding $A_{\textrm{Dop}}$ to \lcdm\ when fitting \planck\ TT results in a $<0.5 \sigma$ shift of the posterior distribution of $A_{\textrm{Dop}}$ from the fiducial value of unity. From a phenomenological perspective, these tests provide no significant evidence for an improvement over \lcdm\ by solely modifying the monopole or dipole contributions to the CMB photon distribution.

The \lcdm\ + $A_{\textrm{Pol}}$ fit to \planck\ TT generally results in no substantial shifts in the central value of the posteriors. The largest such shift is a 0.41 km$\textrm{s}^{-1}\textrm{Mpc}^{-1}$ shift upward in the mean value of $H_0$. Nevertheless there are substantial increases in the uncertainties of the parameters over the \lcdm\ case. In particular, note that the uncertainties of $H_0$ and $\omega_b$ increase by roughly 35$\%$ and 50$\%$ respectively over the \lcdm\ case. This highlights the importance of the polarization effect even when determining parameters from the TT spectrum. 

\begin{figure*}[!tbp]
  \begin{minipage}[b]{0.4\textwidth}
    \hspace*{0cm}
    \includegraphics[width=6.5in, height=6.5in]{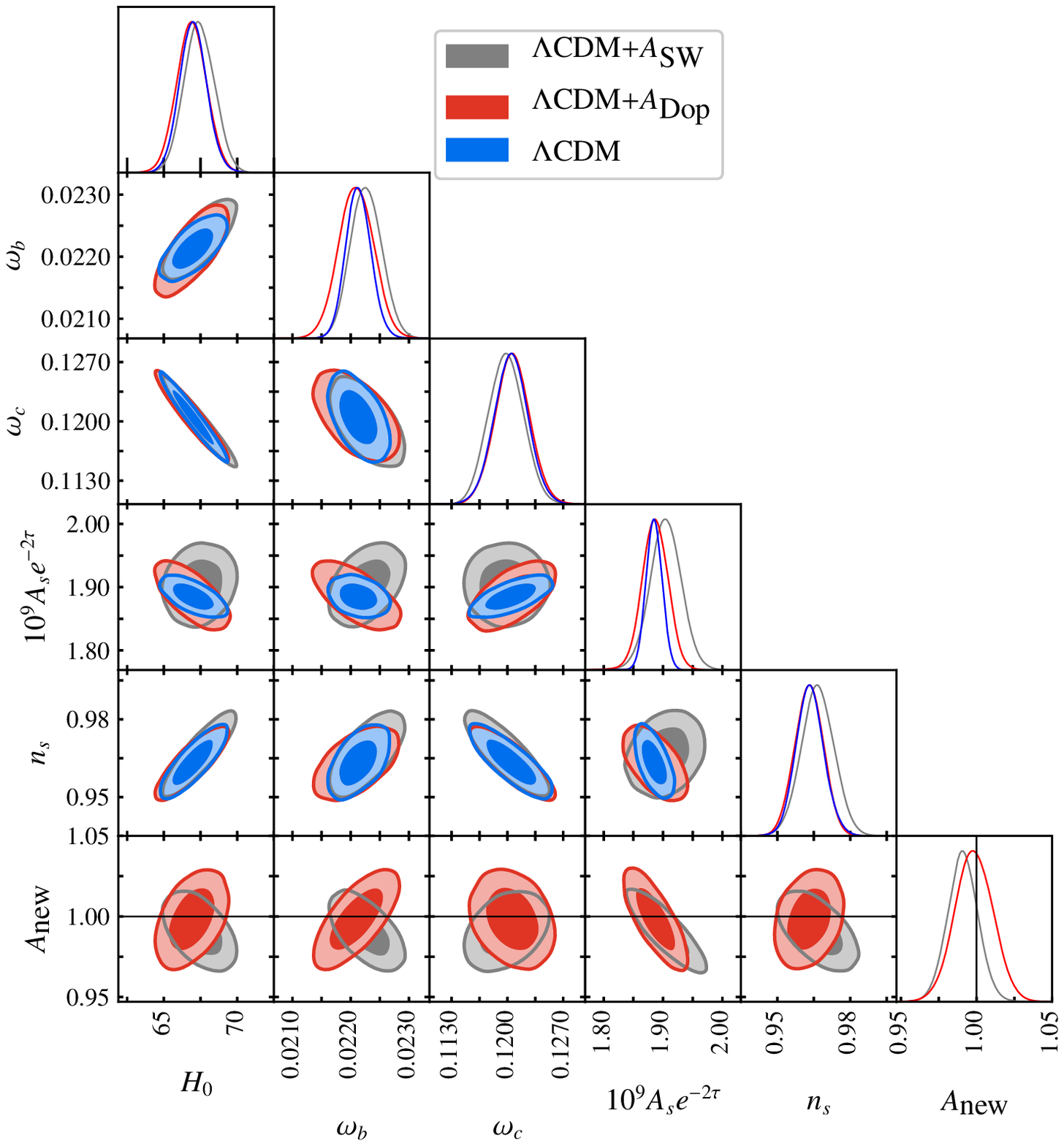}
  \end{minipage}
  \caption{Posterior comparisons of \lcdm\ vs \lcdm\ $+ A_{\textrm{SW}}$ and \lcdm\ $+ A_{\textrm{Dop}}$ fits to \planck\ TT are shown. Neither phenomenological amplitude results in a significant increase in the preferred value of $H_0$ nor a deviation of the phenomenological amplitude from the fiducial value of 1. For $\omega_b$, $\omega_c$, and $n_s$, varying the phenomenological amplitudes keeps the \lcdm\ degeneracy directions intact, but there is a clear change to the degeneracy directions involving $A_se^{-2\tau}$. There is a difference in the sign of the degeneracy direction between $A_{\textrm{SW}}$ and $A_{\textrm{Dop}}$ and the cosmological parameters. This is because increasing $A_{\textrm{SW}}$ disproportionately adds more power to the odd peaks and $A_{\textrm{Dop}}$ disproportionately adds more power to the even peaks. In all cases a prior of $\tau = 0.0506 \pm 0.0086$ was used.
 \label{fig:Figsw_dop}}
  \end{figure*}

In Figure 2 we compare the two dimensional posterior distributions for \lcdm\ + either $A_{\textrm{SW}}$ or $A_{\textrm{Dop}}$ to the \lcdm\ case. The correlations between either $A_{\textrm{SW}}$ or $A_{\textrm{Dop}}$ and the \lcdm\ parameters have an opposite sign for these two models because these phenomenological amplitudes disproportionately add power to either odd or even acoustic peaks of the power spectrum as discussed in Section 2. For example, increasing $A_{\textrm{SW}}$ disproportionately adds power to the odd peaks which must then be compensated by decreasing the baryon density. In contrast, increasing $A_{\textrm{Dop}}$ disproportionately adds power to even peaks which must then be compensated for by increasing the baryon density. In Figure 3, we show the constraints for \lcdm\ and \lcdm\ + one of $A_L$, $A_{\textrm{eISW}}$, or $A_{\textrm{Pol}}$. In all of these cases, the size of the contours increase dramatically over \lcdm, which should be contrasted with the relatively minor changes when varying either $A_{\textrm{SW}}$ or $A_{\textrm{Dop}}$. 

\begin{figure*}[!tbp]
  \begin{minipage}[b]{0.4\textwidth}
    \hspace*{0cm}
    \includegraphics[width=6.5in, height=6.5in]{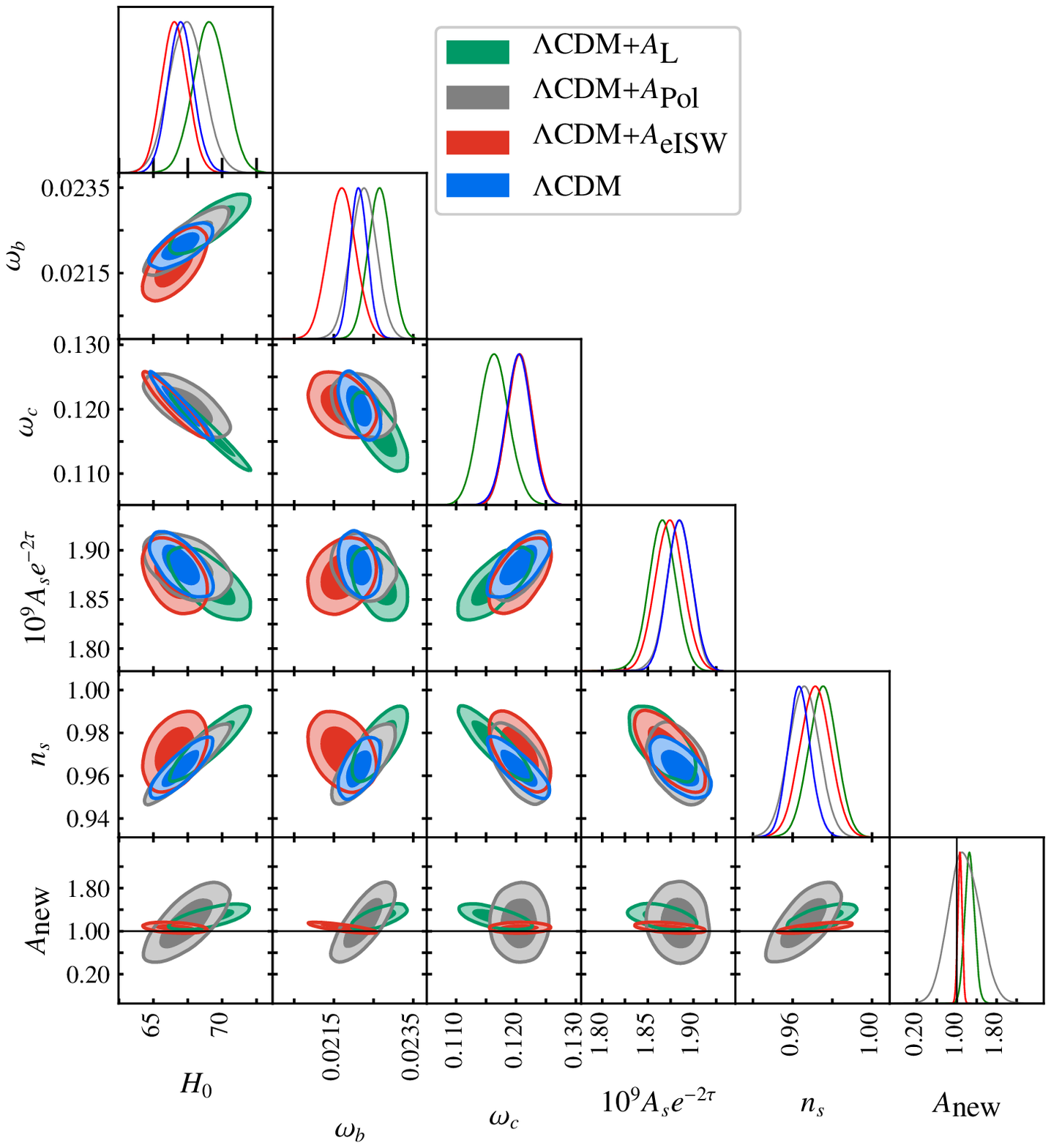}
  \end{minipage}
  \caption{Posterior comparisons of \lcdm\ vs \lcdm\ $+A_{\textrm{L}}$, \lcdm\ $+A_{\textrm{eISW}}$, and \lcdm\ $+A_{\textrm{Pol}}$ fits to \planck\ TT. The \planck\ TT data prefer $A_{\textrm{eISW}} > 1$ at $>1\sigma$ and $A_{\textrm{L}} > 1$ at $>2\sigma$. Note that varying either of these two phenomenological amplitudes tends to shift the baryon density, $\omega_b$, in opposite directions as seen by the orthogonal degeneracy directions. Varying $A_{\textrm{Pol}}$ results in significant degradation in the overall \lcdm\ parameter precision which can seen most clearly in the greater than 30$\%$ increase in the error in $H_0$ and the 50$\%$ increase in the error on $\omega_b$.  In all cases a prior of $\tau = 0.0506 \pm 0.0086$ was used.
 \label{fig:Fig3}}
  \end{figure*}

In summary, these tests show that \lcdm\ is able to correctly account for the Sachs-Wolfe, eISW, Doppler, and Polarization effects measured by \planck\ with the known caveat that there is an internal tension in the \planck\ data between low $\ell$ and high $\ell$, which can be relieved by allowing a parameter like $A_L$ to vary. Because the cosmological parameters do not shift much when the amplitudes for the Sachs-Wolfe, Doppler, eISW, or Polarization effects are varied, the parameter constraints from these physical processes are internally consistent. Finally we note that even when allowing the amplitudes for any one of the physical effects that source the CMB temperature anisotropy to vary, \planck\ TT is still able to place strong constraints on the \lcdm\ parameters.

\begin{deluxetable*}{ccccccc}
 \centering
\tablewidth{6.5in}
 \tablecaption{Mean values and 68$\%$ credible intervals \lcdm\ plus one or more phenomenological amplitudes for the MCMC chains fit to \planck\ 2018 TT Full $\ell$. For definitions of the phenomenological amplitudes see Section 2. We use a prior of $\tau = 0.0506 \pm 0.0086$.  \label{tab:best-fit-multi}}
\tablehead{
 \colhead{Parameter} & \colhead{$+A_L$} & \colhead{$+A_{\textrm{SW}}$} & \colhead{$+A_{\textrm{Dop}}$} & \colhead{$+A_{\textrm{eISW}}$} &  \colhead{$+A_{\textrm{SW}} + A_{\textrm{Dop}}$} & \colhead{$+A_{\textrm{SW}} + A_{\textrm{Dop}} +A_{\textrm{eISW}} $} 
 }
 \startdata
 $H_0$  & 69.11 $\pm$ 1.20 & 67.41 $\pm$ 1.03 & 66.90 $\pm$ 1.02 &  66.55 $\pm$ 0.96 &67.52 $\pm$ 1.04 & 68.59 $\pm$ 1.46  \\
 $100*\omega_b$  & 2.265 $\pm$ 0.029 & 2.226 $\pm$ 0.027 & 2.210 $\pm$ 0.030 & 2.170 $\pm$ 0.035  & 2.180 $\pm$ 0.032 & 2.225 $\pm$ 0.054 \\
 $\omega_c$ & 0.1164 $\pm$ 0.0025 & 0.1198 $\pm$ 0.0022 & 0.1206 $\pm$ 0.0022 & 0.1206 $\pm$ 0.0021 &0.1187 $\pm$ 0.0023 & 0.1173 $\pm$ 0.0026  \\
 $10^9A_se^{-2\tau}$ & 1.8658 $\pm$ 0.0156 & 1.904 $\pm$ 0.027 & 1.886 $\pm$ 0.022 & 1.873 $\pm$ 0.0165 & 2.145 $\pm$ 0.107 & 2.32 $\pm$ 0.20  \\
 $n_s$ & 0.9751 $\pm$ 0.0072 & 0.9666 $\pm$ 0.0067 & 0.9631 $\pm$ 0.0058 & 0.9713 $\pm$ 0.0077 & 0.9786 $\pm$ 0.0084 & 0.9783 $\pm$ 0.0085\\
 $A_L$ & 1.259 $\pm$ 0.099 & --------- & --------- & --------- & --------- & --------- \\
 $A_{\textrm{SW}}$ & --------- & 0.9909 $\pm$ 0.0100 & --------- & --------- & 0.936 $\pm$ 0.023 & 0.903 $\pm$ 0.039 \\
 $A_{\textrm{Dop}}$ & --------- &--------- & 0.9984 $\pm$ 0.0126 & --------- & 0.925 $\pm$ 0.029 & 0.890 $\pm$ 0.043 \\
 $A_{\textrm{eISW}}$ & --------- & --------- & --------- & 1.064 $\pm$ 0.042 & --------- & 0.929 $\pm$ 0.067 \\
 \hline
 $\Omega_mh^2$  &  0.1390 $\pm$ 0.0023 &  0.1421 $\pm$ 0.0020 & 0.1427 $\pm$ 0.0020 & 0.1423 $\pm$ 0.0020 & 0.1405 $\pm$ 0.0022 & 0.1395 $\pm$ 0.0023 \\
 $\Omega_m$  & 0.2915 $\pm$ 0.0148 & 0.3129 $\pm$ 0.0141 & 0.3186 $\pm$ 0.0139 & 0.3217 $\pm$ 0.0134  & 0.3085 $\pm$ 0.0140 & 0.2972 $\pm$ 0.0172 \\
 $\sigma_8$ & 0.7933 $\pm$ 0.0120 & 0.8147 $\pm$ 0.0099 & 0.8136 $\pm$ 0.0114 & 0.8144 $\pm$ 0.00978  & 0.865 $\pm$ 0.023 & 0.891 $\pm$ 0.033\\
 \hline
 $\chi^2$ & 221.45 & 228.59 & 229.49& 227.24& 222.53 & 220.33 \\
 $\chi^2_{\Lambda CDM} -\chi^2$ & 8.5 & 0.91 & 0.01 & 2.26 & 6.97 & 9.17
 \enddata
 \vspace{-0.5cm}
\end{deluxetable*}

\subsection{\lcdm\ $+ A_{\textrm{SW}} + A_{\textrm{Dop}} $}
In the previous subsection, we showed results for extending \lcdm\ to include one of the phenomenological amplitudes that we introduced in Section 2. In this subsection, we discuss adding pairs of the phenomenological amplitudes. In general, we find that much like adding one phenomenological amplitude, adding pairs of phenomenological amplitudes does not result in either an improved fit to \planck\ TT or a reduction in the $H_0$ tension with late universe measurements. We find that only \lcdm\ + $A_{\textrm{SW}}$ + $A_{\textrm{Dop}}$ exhibits a significant improvement to the fit to \planck\ TT.

We summarize the results of the MCMC sampling for \lcdm\ + $A_{\textrm{SW}}$ + $A_{\textrm{Dop}}$ to \planck\ TT in Table 2. With two parameters, it becomes more complicated to define when there is a significant shift in the posterior, but $A_{\textrm{SW}}$ and $A_{\textrm{Dop}}$ are both more than 2$\sigma$ below the fiducial value of unity when simultaneously allowed to vary. Additionally, the PTE of the $\Delta \chi^2$ assuming two degrees of freedom is 0.03 indicating a significant improvement over the \lcdm\ case. Note that adding both $A_{\textrm{SW}}$ and $A_{\textrm{Dop}}$ together results in a significant improvement over standard \lcdm\ when fitting to \planck\ TT because when only one at a time was added there was much less improvement. Allowing both $A_{\textrm{SW}}$ and $A_{\textrm{Dop}}$ to vary simultaneously does not also increase the \planck\ TT preferred value of $H_0$ like when adding $A_L$.

Note that since $A_{\textrm{SW}}$ and $A_{\textrm{Dop}}$ appear to be acting in unison, we should recover approximately the same model if we use a single phenomenological amplitude to scale both the Sachs-Wolfe and Doppler effects. Taking a step back, if we had used a single amplitude to rescale all of the effects that source the CMB TT anisotropy in Equation 2, then this new phenomenological amplitude would have been almost completely degenerate with $A_s$, up to corrections from lensing, when fitting to \planck\ TT. In this case, $A_s$ becomes a proxy for $A_L$ because of how $A_s$ explicitly enters the equations for lensing \citep[see, e.g., Section 3.1-3.2 of][]{lewis/challinor:2006}. Varying both $A_{\textrm{SW}}$ and $A_{\textrm{Dop}}$ simultaneously increases the uncertainty of $A_se^{-2\tau}$ by a factor of 4 relative to the \lcdm\ case, which allows sufficient freedom for $A_s$ to become a proxy for $A_L$. 

Additionally in Table 2, we include the constraints when $A_{\textrm{eISW}}$ is added to \lcdm\ + $A_{\textrm{SW}}$ + $A_{\textrm{Dop}}$. For this MCMC run, we only used a convergence criteria of $R-1 = 0.1$ because convergence was difficult to achieve. While this \lcdm\ + $A_{\textrm{SW}}$ + $A_{\textrm{Dop}}$ + $A_{\textrm{eISW}}$ result gives a significant improvement over \lcdm\ with a PTE of 0.03 assuming a $\Delta \chi^2$ with three degrees of freedom, it is not a significant improvement over \lcdm\ + $A_{\textrm{SW}}$ + $A_{\textrm{Dop}}$ with a PTE of 0.13 assuming one degree of freedom. This improved $\Delta \chi^2$ is roughly equivalent to the improved $\Delta \chi^2$ when adding only $A_{\textrm{eISW}}$ to \lcdm\ as shown in Table 1, but note that \planck\ TT prefers $A_{\textrm{eISW}} < 1$ for this model to be more in line with the preferred values for $A_{\textrm{SW}}$ and $A_{\textrm{Dop}}$. While adding $A_{\textrm{eISW}}$ to \lcdm\ + $A_{\textrm{SW}}$ + $A_{\textrm{Dop}}$ does not result in a significant improvement, there is an increase in the preferred value of $H_0$ similar to the \lcdm\ + $A_L$ preferred $H_0$ value.

In Figure 4, we compare the 2D posteriors for the one parameter extensions, \lcdm\ + $A_{\textrm{SW}}$, $A_{\textrm{Dop}}$, and $A_{\textrm{eISW}}$, and the combinations \lcdm\ + $A_{\textrm{SW}}$ + $A_{\textrm{Dop}}$ and \lcdm\ + $A_{\textrm{SW}}$ + $A_{\textrm{Dop}}$ + $A_{\textrm{eISW}}$. Note the strong degeneracies between the phenomenological amplitudes and the scalar amplitude when more than one phenomenological amplitude is varied. Adding $A_{\textrm{eISW}}$ results in a substantial increase in the degeneracy between the phenomenological amplitudes and the scalar amplitude. This, in turn, allows for parameters like $H_0$ to access a broader parameter space. 

\begin{figure*}[!tbp]
  \begin{minipage}[b]{0.4\textwidth}
    \hspace*{-1.6cm}
    \includegraphics[width=7.5in, height=7.5in]{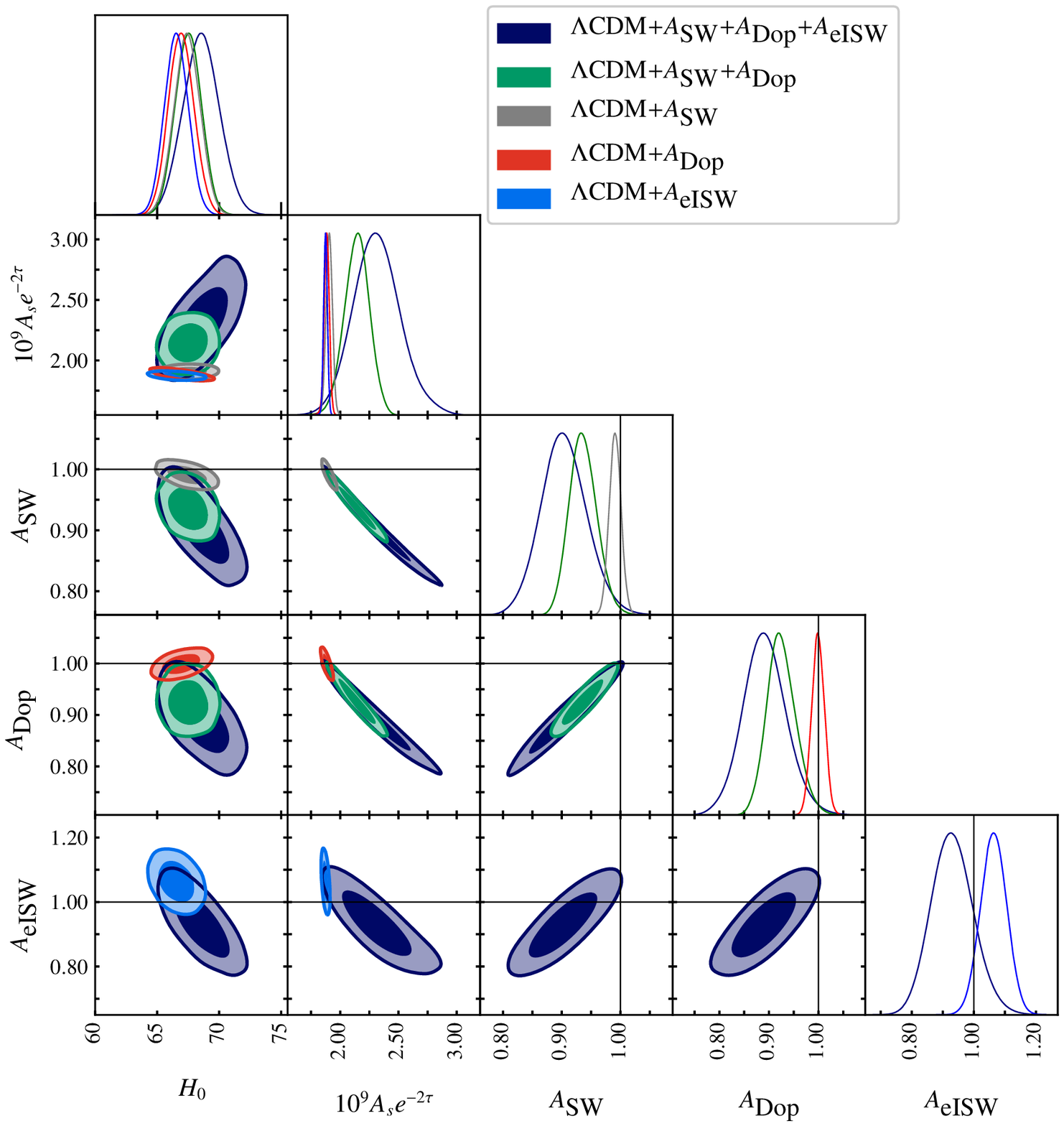}
  \end{minipage}
  \caption{ Posteriors of \lcdm\ $+ A_{\textrm{SW}} + A_{\textrm{Dop}} + A_{\textrm{eISW}}$, \lcdm\  $+ A_{\textrm{SW}} + A_{\textrm{Dop}}$, \lcdm\ $+ A_{\textrm{SW}}$, \lcdm\ $+ A_{\textrm{Dop}}$, and \lcdm\ $+ A_{\textrm{eISW}}$  fits to \planck\ TT. Including both $A_{\textrm{SW}}$ and $A_{\textrm{Dop}}$ results in a strong degeneracy between the phenomenological amplitudes and $A_se^{-2\tau}$ which was not necessarily expected given the posteriors when only one of them is varied. The model \lcdm\ $+ A_{\textrm{SW}} + A_{\textrm{Dop}}$ prefers values for $A_{\textrm{SW}}$ and $A_{\textrm{Dop}}$ that deviate from unity by about 2$\sigma$. Additionally including $A_{\textrm{eISW}}$ results in an even stronger degeneracy between the phenomenological amplitudes and $A_s$, which notably broadens the allowed parameter space for parameters like $H_0$. Nevertheless, the result is still consistent with $A_{\textrm{eISW}} = 1$, which suggests that adding $A_{\textrm{eISW}}$ does not result in a significant improvement over \lcdm\ $+ A_{\textrm{SW}} + A_{\textrm{Dop}}$. In all cases there is a prior of $\tau = 0.0506 \pm 0.0086$.
 \label{fig:Figsw_dop_eisw}}
  \end{figure*}

In Figure 5, we demonstrate how the Sachs-Wolfe and Doppler effects work together to rescale the power spectrum by plotting the quotient of the \lcdm\ + $A_{\textrm{SW}}$ + $A_{\textrm{Dop}}$ case to the \lcdm\ case. In particular note that for $\ell > 400$, the quotient is flat, up to some wiggles that result from not additionally rescaling $A_{\textrm{Pol}}$. In the middle panel of Figure 5, we show that the slope in the quotient for $\ell < 400$ results from not also rescaling the ISW effect. For \lcdm\ + $A_{\textrm{SW}}$ + $A_{\textrm{Dop}}$, it is this ability to rescale the TT power spectrum on scales $\ell > 400$ that degrades the precision of $A_s$ allowing it to become a proxy for $A_L$. 

\begin{figure}[!tbp]
  \centering
  \begin{minipage}[b]{0.4\textwidth}
    \hspace*{-1.6cm}
    \includegraphics[width=4in, height=7.5in]{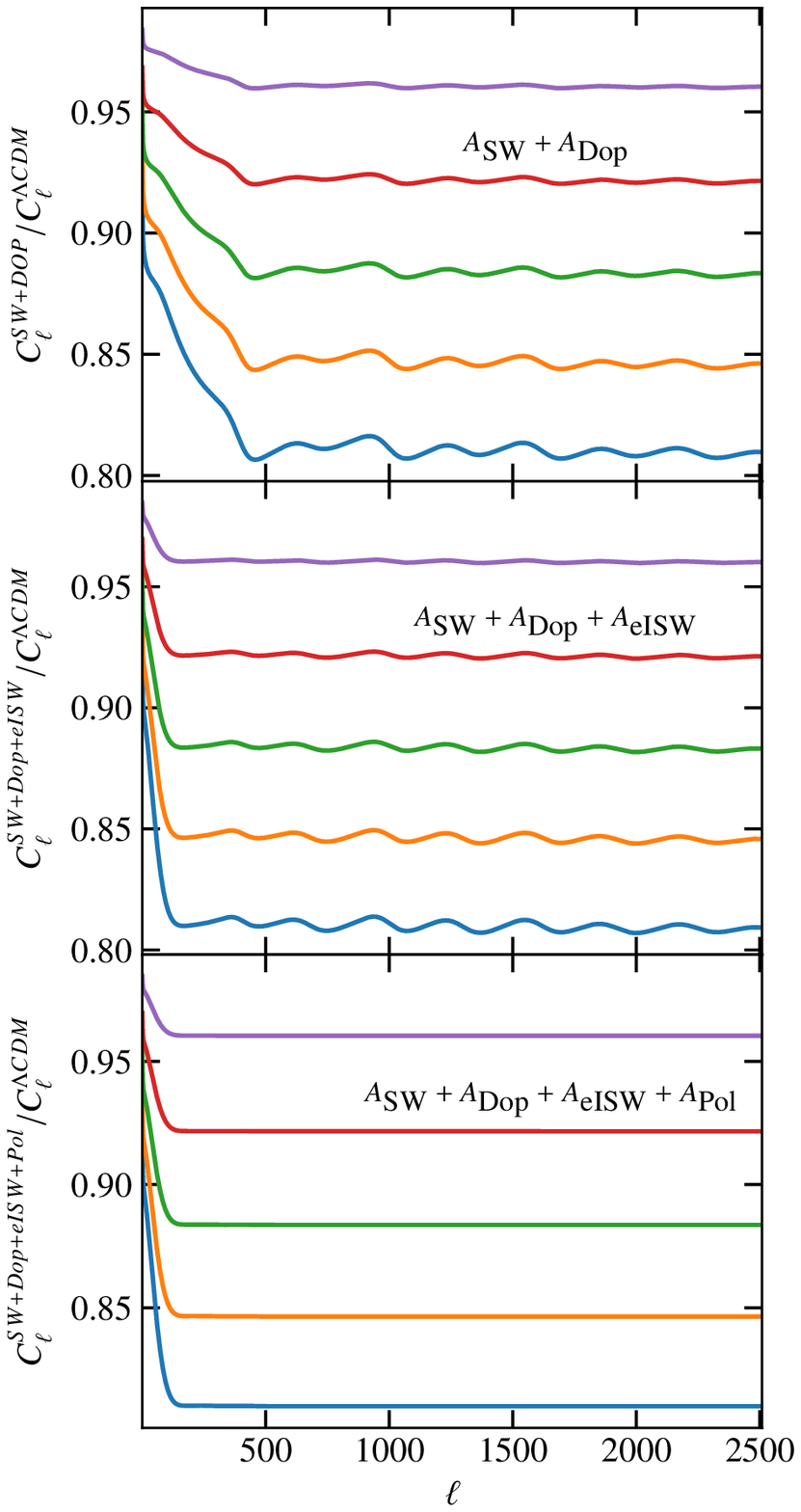}
  \end{minipage}
  \caption{Top panel: ratio of the TT power spectrum when varying both $A_{\textrm{SW}}$ and $A_{\textrm{Dop}}$ together to the TT power spectrum of standard \lcdm. The curves correspond to all phenomenological amplitudes being set to $\{0.90,0.92,0.94,0.96,0.98\}$ from top to bottom. If $A_{\textrm{SW}}$ and $A_{\textrm{Dop}}$ act in unison, then, at high $\ell$, they can rescale the power spectrum and allow $A_s$ to become a proxy for $A_L$, which results in a significant improvement over \lcdm\ by mimicking the effect of lensing. Middle panel: when $A_{\textrm{eISW}}$ acts in unison with $A_{\textrm{SW}}$ and $A_{\textrm{Dop}}$ the degeneracy extends to lower multipole moments resulting in a stronger degeneracy. Bottom panel: the wiggles at high $\ell$ result from not including $A_{\textrm{Pol}}$ in the rescaling. 
 \label{fig:multiparamamps}}
\end{figure}

Because the degeneracy between $A_{\textrm{SW}}$, $A_{\textrm{Dop}}$, and $A_s$ when fitting \lcdm\ + $A_{\textrm{SW}}$ + $A_{\textrm{Dop}}$ breaks down for multipoles $\ell < 400$, we use fits to \planck\ TT and \planck\ TT $\ell > 800$ to illustrate that \lcdm\ + $A_{\textrm{SW}}$ + $A_{\textrm{Dop}}$ is approximately finding the \lcdm\ + $A_L$ solution. In Figure 6, we show the residuals of the theory TT power spectrum calculated using the best-fit parameters for the \lcdm\ + $A_L$ and \lcdm\ + $A_{\textrm{SW}}$ + $A_{\textrm{Dop}}$ fits to both \planck\ TT and \planck\ TT $\ell > 800$ with the theory TT power spectrum calculated using the best-fit parameters for the \lcdm\ fit to \planck\ TT. Additionally, we include the residual of the measured \planck\ TT data with the \lcdm\ fit to \planck\ TT. To increase the clarity of the plot, we rebin the \planck\ TT data using new super bins of $\Delta \ell \approx 65$. Note that there are high levels of correlation, often at the 80$\%$ level, between the bins for Plik Lite which result from marginalizing over the foregrounds. 
  
\begin{figure*}[!tbp]
 \centering
    \hspace*{-0.75cm}
    \includegraphics[width=7in, height=5in]{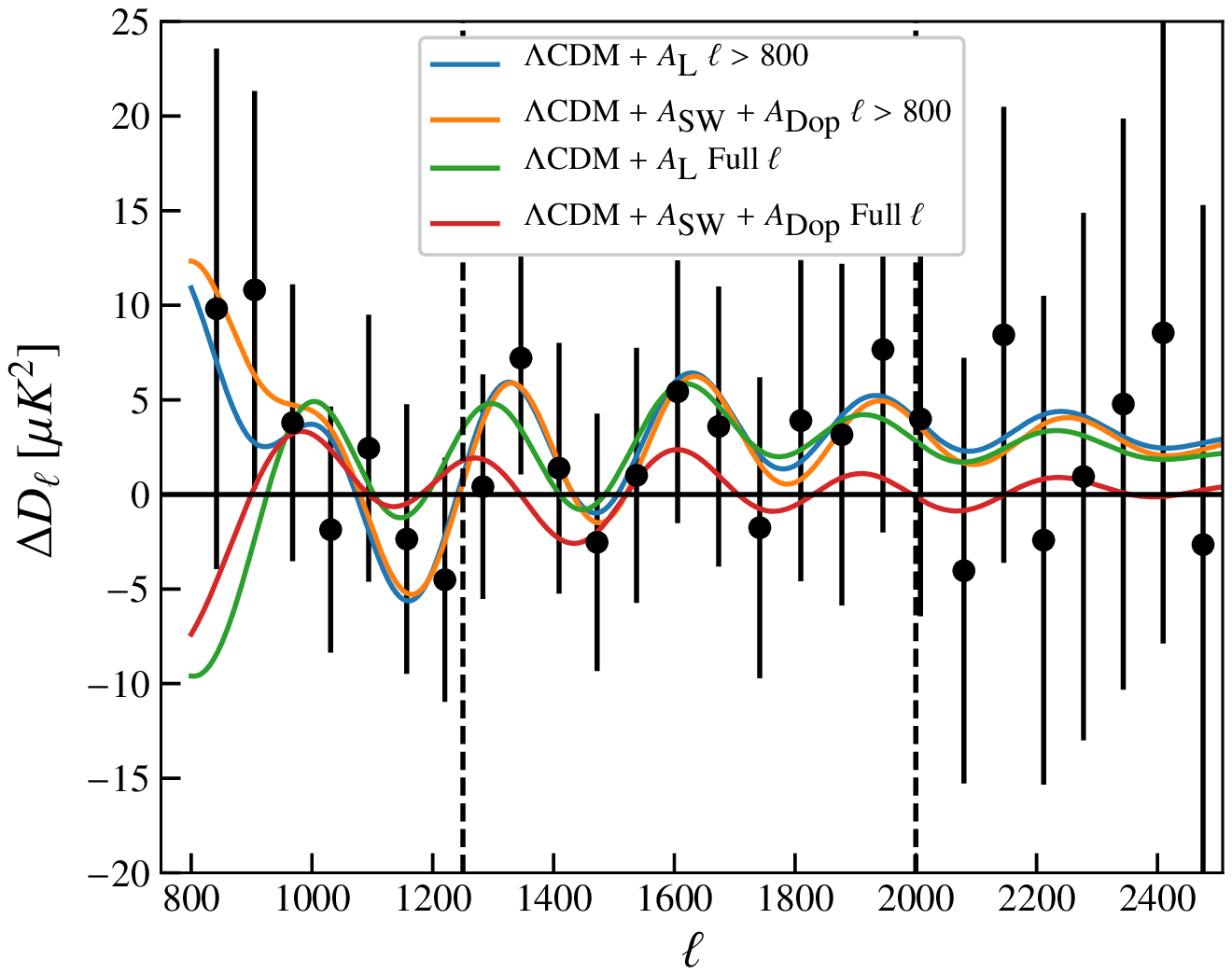}
  \caption{Residuals of the best-fit TT power spectrum from \lcdm\ + $A_L$ and \lcdm\ + $A_{\textrm{SW}}$ + $A_{\textrm{Dop}}$ fits to both \planck\ TT and \planck\ TT $\ell > 800$ with the \lcdm\ fit to \planck\ TT. We additionally include the residual of the measured \planck\ TT data (black) points, but we have rebinned them with $\Delta \ell \approx 65$ for visual clarity. When fit to only \planck\ TT $\ell > 800$, varying the amplitudes for the Sachs-Wolfe and Doppler effects results in a similar residual to the \lcdm\ + $A_L$ residual, which suggests that at high $\ell$ these two models achieve approximately the same effect. The \lcdm\ + $A_{\textrm{SW}}$ + $A_{\textrm{Dop}}$ fit to \planck\ TT is restricted predominantly by the ISW effect breaking the degeneracy between the $A_{\textrm{SW}}$, $A_{\textrm{Dop}}$, and $A_s$, but it still fits the oscillatory residual in the multipole range $1250 \leq \ell \leq 2000$, which explains the improved fit to the $\chi^2$ over the \lcdm\ fit to \planck\ TT. Note that the bins provided by the \planck\ collaboration for Plik Lite are highly correlated at high $\ell$.
 \label{fig:residual}}
  \end{figure*}

Importantly, Figure 6 shows that the residuals for \lcdm\ + $A_L$ and \lcdm\ + $A_{\textrm{SW}}$ + $A_{\textrm{Dop}}$ are highly correlated when fit to \planck\ TT $\ell > 800$, which emphasizes that these two models are making the same changes at high $\ell$, and it is the low $\ell$ behavior that restricts the latter model when fit to \planck\ TT. For $\ell > 1250$, the \lcdm\ + $A_L$ fit to \planck\ TT also becomes highly correlated to these fits indicating that this is the primary feature of the lensing solution. Further note that the \lcdm\ + $A_{\textrm{SW}}$ + $A_{\textrm{Dop}}$ fit to \planck\ TT does fit the oscillatory residual in the \planck\ data, albeit without the increased power for multipoles $\ell > 1250$. This is how the \lcdm\ + $A_{\textrm{SW}}$ + $A_{\textrm{Dop}}$ fit to \planck\ TT achieves a significant improvement over \lcdm. 
  
In Table 3, we show the results from MCMC runs for \lcdm, \lcdm\ + $A_L$, and \lcdm\ + $A_{\textrm{SW}}$ + $A_{\textrm{Dop}}$ fits to \planck\ TT $\ell \leq 800$ and \planck\ TT $\ell > 800$. Neither \lcdm\ + $A_L$ nor \lcdm\ + $A_{\textrm{SW}}$ + $A_{\textrm{Dop}}$ results in a significantly better fit to the \planck\ temperature data when only half of the data are included. This highlights that the improvement found when allowing either $A_L$ or $A_{\textrm{SW}}$ and $A_{\textrm{Dop}}$ to vary is primarily in bringing the two halves of the \planck\ power spectrum into better agreement. 

Allowing $A_{\textrm{SW}}$ and $A_{\textrm{Dop}}$ to vary when fitting to either \planck\ TT $\ell \leq 800$ or \planck\ TT $\ell > 800$ results in an increase in the preferred value of $H_0$, though notably the uncertainty of $H_0$ also increases to be  $>5$ km$\textrm{s}^{-1}\textrm{Mpc}^{-1}$. For \planck\ TT $\ell > 800$, the uncertainty on $H_0$ increases by a factor of 3.7 when $A_L$ is is allowed to vary indicating that lensing is important in constraining cosmological parameters at high $\ell$. 

\begin{deluxetable*}{ccccccc}
 \centering
\tablewidth{6.5in}

 \tablecaption{Mean values and 68$\%$ credible intervals for \lcdm\, \lcdm\ $+A_L$, and \lcdm\ $ + A_{\textrm{SW}}+A_{\textrm{Dop}}$ MCMC chains fits to \planck\ 2018 TT $\ell \leq 800$ and $\ell > 800$. For definitions of the phenomenological amplitudes see Section 2. We use a prior of $\tau = 0.0506 \pm 0.0086$.  \label{tab:best-fit-data_splits}}

\tablehead{
 \colhead{Parameter} & \colhead{$\Lambda CDM$ $\ell \leq 800$} & \colhead{$A_L$ $\ell \leq 800$} & \colhead{$A_{\textrm{SW}}+A_{\textrm{Dop}}$ $\ell \leq 800$} & \colhead{\lcdm\ $\ell > 800$} & \colhead{$A_L$ $\ell > 800$} & \colhead{$A_{\textrm{SW}}+A_{\textrm{Dop}}$ $\ell > 800$} 
 }
 
 \startdata
 $H_0$ & 69.95 $\pm$ 1.84 & 71.1 $\pm$ 2.1 & 73.3 $\pm$ 5.0  &  64.28 $\pm$ 1.33 & 71.0 $\pm$ 4.8 &  69.93 $\pm$ 5.3 \\
 $100*\omega_b$ & 2.252 $\pm$ 0.042 & 2.283 $\pm$ 0.050 & 2.345 $\pm$ 0.111 & 2.193 $\pm$ 0.039 & 2.387 $\pm$ 0.145 & 2.273 $\pm$ 0.145 \\
 $\omega_c$ & 0.1145 $\pm$ 0.0033 & 0.1128 $\pm$ 0.0036 & 0.1106 $\pm$ 0.0068 & 0.1279 $\pm$ 0.0034 & 0.1152 $\pm$ 0.0088 & 0.1162 $\pm$ 0.0106 \\
 $10^9A_se^{-2\tau}$ & 1.8559 $\pm$ 0.0170 & 1.8477 $\pm$ 0.0177 & 1.80 $\pm$ 0.31 & 1.922 $\pm$ 0.021 & 1.896 $\pm$ 0.025 & 2.59 $\pm$ 0.62\\
 $n_s$ & 0.9756 $\pm$ 0.0120 & 0.9829 $\pm$ 0.0137 & 0.9895 $\pm$ 0.0186 & 0.9489 $\pm$ 0.0119 & 0.9579 $\pm$ 0.0134 & 0.9823 $\pm$ 0.037\\
 $A_{L}$ & ---------& 1.64 $\pm$ 0.53 & --------- & --------- & 1.41 $\pm$ 0.30 & --------- \\
 $A_{\textrm{SW}}$ & ---------& --------- & 1.014 $\pm$ 0.073 & --------- & ---------&  0.873 $\pm$ 0.115 \\
 $A_{\textrm{Dop}}$ & ---------& --------- & 1.045 $\pm$ 0.104 & --------- & ---------&  0.858 $\pm$ 0.142 \\
 \hline
 $\Omega_mh^2$ & 0.1370 $\pm$ 0.0030 & 0.1355 $\pm$ 0.0032 & 0.1341 $\pm$ 0.0053 & 0.1498 $\pm$ 0.0033 & 0.1390 $\pm$ 0.0075 & 0.1389 $\pm$ 0.0099\\
 $\Omega_m$ & 0.281 $\pm$ 0.021 & 0.269 $\pm$ 0.022 & 0.255 $\pm$ 0.049 & 0.363 $\pm$ 0.023 & 0.282 $\pm$ 0.053& 0.292 $\pm$ 0.067\\
 $\sigma_8$ & 0.7858 $\pm$ 0.0139  & 0.7789 $\pm$ 0.0150 & 0.758 $\pm$ 0.087 & 0.8371 $\pm$ 0.0124 & 0.783 $\pm$ 0.040 & 0.923 $\pm$ 0.086 \\
 \hline
 $\chi^2$ & 95.48 & 94.04 & 94.35 & 123.28 & 121.10 & 121.32
 \enddata
 \vspace{-0.5cm}
\end{deluxetable*}

In summary, we find that adding the phenomenological amplitudes we introduced in Section 2 in pairs does not result in a significant improvement to the fit to \planck\ TT over standard \lcdm.  The one exception is when the phenomenological amplitudes for the Sachs-Wolfe and Doppler effects are both allowed to vary, but we show that this solution is really approximately refinding the \lcdm\ + $A_L$ solution by allowing $A_s$ to become a proxy for $A_L$. Adding more phenomenological amplitudes, such as $A_{\textrm{eISW}}$, can make this approximation marginally better but does not result in a significant improvement to the fit to \planck\ TT. 

\section{Can Additional Model Freedom Improve Over \lcdm\ + $A_L$?}
In the previous section we found that none of the phenomenological amplitudes that we introduced in Section 2 showed any significant deviations from standard \lcdm\ predictions. Moreover, while we found that combining the phenomenological amplitudes for the Sachs-Wolfe and Doppler effects show a $\sim2.7\sigma$ preference for $A_{\textrm{SW}}$ and $A_{\textrm{Dop}}$ both below unity, we noted that this solution was just refinding the \lcdm\ + $A_L$ solution. 

In this section we test some extensions to \lcdm\ + $A_L$ to seek a model extension that better fits \planck\ TT. In Section 4.1, we test extending the \lcdm\ + $A_L$ model to include the phenomenological amplitudes that we introduced in Section 2. In Section 4.2, we test extending the \lcdm\ + $A_L$ model to include one of $N_{\textrm{eff}}$, $n_{\textrm{run}}$, and $Y_{\textrm{He}}$, which all have effects on the high multipole moments of the TT power spectrum. 

\subsection{Testing $A_L$ Plus One Additional Phenomenological Amplitude}

In this subsection, we add the phenomenological amplitudes introduced in Section 2 to \lcdm\ + $A_L$ and fit to \planck\ TT. While the results in Section 3.1 showed no preference for any of these phenomenological amplitudes alone, Section 3.2 highlights the possibility that multiple phenomenological amplitudes working together could result in some improvement to the fit to \planck\ TT. 

\begin{deluxetable*}{cccccc}
 \centering
\tablewidth{6.5in}

 \tablecaption{Mean values and 68$\%$ credible intervals for  \lcdm\ $+A_L$ plus one phenomenological parameter for the MCMC chains fit to \planck\ 2018 TT Full $\ell$. For definitions of the phenomenological amplitudes see Section 2. We use a prior of $\tau = 0.0506 \pm 0.0086$.  \label{tab:best-fit-lens+}}

\tablehead{
 \colhead{Parameter} & \colhead{$+ A_L$} & \colhead{$+ A_L + A_{\textrm{SW}}$} & \colhead{$+ A_L + A_{\textrm{Dop}}$} & \colhead{$+ A_L + A_{\textrm{eISW}}$} & \colhead{$+ A_L + A_{\textrm{Pol}}$} 
 }
 
 \startdata
 $H_0$ & 69.11 $\pm$ 1.20 & 69.04 $\pm$ 1.20 & 68.87 $\pm$ 1.23 & 69.02 $\pm$ 1.39 &  69.45 $\pm$ 1.55 \\
 $100*\omega_b$  & 2.265 $\pm$ 0.029 & 2.259 $\pm$ 0.030 & 2.248 $\pm$ 0.035 & 2.258 $\pm$ 0.051 & 2.274 $\pm$ 0.040\\
 $\omega_c$ & 0.1164 $\pm$ 0.0025 & 0.1162 $\pm$ 0.0025 & 0.1165 $\pm$ 0.0025 & 0.1164 $\pm$ 0.0026 & 0.1165 $\pm$ 0.0025 \\
 $10^9A_se^{-2\tau}$  & 1.8658 $\pm$ 0.0156 & 1.838 $\pm$ 0.036 & 1.880 $\pm$ 0.024 & 1.8643 $\pm$ 0.0169 & 1.8635 $\pm$ 0.0176\\
 $n_s$  & 0.9751 $\pm$ 0.0072 & 0.9736 $\pm$ 0.0073 & 0.9748 $\pm$ 0.0072 & 0.9761 $\pm$ 0.0080 & 0.9768 $\pm$ 0.0090\\
 $A_L$ & 1.259 $\pm$ 0.099 & 1.329 $\pm$ 0.126 & 1.286 $\pm$ 0.105 & 1.258 $\pm$ 0.111 &  1.263 $\pm$ 0.100 \\
 $A_{new}$ & --------- & 1.0116 $\pm$ 0.0128 & 0.9874 $\pm$ 0.0137 & 1.009 $\pm$ 0.047 &  1.13 $\pm$ 0.36 \\
 \hline
 $\Omega_mh^2$  & 0.1390 $\pm$ 0.0023 & 0.1388 $\pm$ 0.0023 & 0.1389 $\pm$ 0.0023 & 0.1390 $\pm$ 0.0023 & 0.1392 $\pm$ 0.0023 \\
 $\Omega_m$  & 0.2915 $\pm$ 0.0148 & 0.2917 $\pm$ 0.0148 & 0.2934 $\pm$ 0.0151 & 0.2923 $\pm$ 0.0164 & 0.2892 $\pm$ 0.0161 \\
 $\sigma_8$  & 0.7933 $\pm$ 0.0120 & 0.7866 $\pm$ 0.0141 & 0.7973 $\pm$ 0.0128 & 0.7938 $\pm$ 0.0129 & 0.7939 $\pm$ 0.0127 \\
 \hline
 $\chi^2$  & 221.45& 220.98 & 220.72 & 221.44 & 221.40 \\
 $\chi^2_{\Lambda CDM + A_L} - \chi^2$  & 0 & 0.47 & 0.73 & 0.01 & 0.05
 \enddata

 \vspace{-0.5cm}

\end{deluxetable*}

The results of adding $A_{\textrm{SW}}$, $A_{\textrm{Dop}}$, $A_{\textrm{eISW}}$, or $A_{\textrm{Pol}}$ to \lcdm\ + $A_L$ when fitting to \planck\ TT are summarized in Table 4. From Table 4, it is clear that there is almost no improvement to the $\chi^2$ when including these phenomenological amplitudes. Moreover, none of the posteriors for the phenomenological amplitudes are more than 1$\sigma$ away from unity. This is consistent with our results from Section 3.1 but again highlights that each of these physical effects are being correctly accounted for. 

In Section 3.1, we showed that \lcdm\ + $A_{\textrm{eISW}}$ results in a minor improvement of 2.26 in the $\chi^2$ fit to \planck\ TT over \lcdm\ alone. Adding $A_{\textrm{eISW}}$ to \lcdm\ + $A_L$ results in almost no change in the $\chi^2$ from the \lcdm\ + $A_L$ case nor a significant shift in the preferred value of $A_{\textrm{eISW}}$ from unity. This suggests that the changes made by varying $A_{\textrm{eISW}}$ are no longer necessary when $A_L$ is already allowed to vary. This is consistent with the primary improvement to the fit to \planck\ TT found in the \lcdm\ + $A_{\textrm{eISW}}$ model coming from multipoles $\ell < 30$ as \lcdm\ + $A_L$ already makes improvements to fitting these multipoles. Note that the improvement in the multipole range $\ell < 30$ when \lcdm\ + $A_L$ is fit to \planck\ TT results from freeing up the constraints on other cosmological parameters such as allowing the preferred value of $A_s$ to decrease and the preferred value of $n_s$ to increase. 

The model \lcdm\ + $A_L$ + $A_{\textrm{eISW}}$ also provides an exploration into how the physical matter density is constrained by the CMB TT power spectrum. \cite{Knox/etal:2020} point out that the physical matter density is predominantly determined by the overall photon envelope, followed by lensing, and then finally by the eISW effect. When $A_L$ is allowed to vary, the uncertainty of the physical matter density increases by roughly 16.5$\%$ over standard \lcdm. Meanwhile, allowing $A_{\textrm{eISW}}$ to vary results in a 3.5$\%$ increase in the uncertainty over standard \lcdm. Allowing $A_L$ and $A_{\textrm{eISW}}$ to vary results in a 19$\%$ increase in the physical matter density over standard \lcdm. This suggests that the overall photon envelope constrains the physical matter density significantly more than either lensing or the eISW effect, consistent with \cite{Knox/etal:2020}. However, note that Table 3 shows that allowing $A_L$ to vary results in a 240$\%$ increase in the uncertainty of the physical matter density when only $\ell > 800$ are included. This highlights the importance of lensing to constraining the physical matter density at high $\ell$. 

Note that \lcdm\ + $A_L$ + $A_{\textrm{Dop}}$ shows the largest reduction in the $\chi^2$ despite the fact that adding $A_{\textrm{Dop}}$ to \lcdm\ resulted in the smallest change to the $\chi^2$. This is accompanied by a 0.92$\sigma$ shift downward in preferred value of $A_{\textrm{Dop}}$, both of which indicate that this is not a significant improvement. The preferred value of $A_{\textrm{SW}}$ for the model \lcdm\ + $A_L$ + $A_{\textrm{SW}}$ shifts upward by about 0.9$\sigma$, comparable to the shift in the preferred value of $A_{\textrm{Dop}}$. Adding either $A_{\textrm{SW}}$ or $A_{\textrm{Dop}}$ to \lcdm\ + $A_L$ results in either more power to the odd peaks or similarly less power to the even peaks. In both cases, the preferred value of $A_L$ is able to increase relative to the preferred value of $A_L$ from \lcdm\ + $A_L$ fit to \planck\ TT data. 

Allowing both $A_L$ and $A_{\textrm{Pol}}$ to vary results in a negligible shift in the central values of the posteriors and a negligible improvement to the $\chi^2$ when fitting to \planck\ TT data. Again, the most significant effect when allowing $A_{\textrm{Pol}}$ to vary is an increase in the uncertainty of parameters such as the 30$\%$ increase in the uncertainty of $H_0$. 
 
\subsection{Testing $A_L$ plus one additional non-phenomenological amplitude}
In this subsection we test models for \lcdm\ + $A_L$ + one of $N_{\textrm{eff}}$, $n_{\textrm{run}}$, and $Y_{\textrm{He}}$. $N_{\textrm{eff}}$ is designed to account for the effective number of relativistic degrees of freedom well after electron-positron annihilation. The parameter $n_{\textrm{run}}$ accounts for possible linear order deviations from a flat primordial power spectrum with a spectral tilt given by $n_s$. Finally, the helium fraction, $Y_{\textrm{He}}$, affects the free electron density before and during recombination. All of these parameters added to \lcdm\ + $A_L$ could in principle affect the low $\ell$ and high $\ell$ consistency. 

\begin{deluxetable*}{ccccc}
 \centering
\tablewidth{6.5in}

 \tablecaption{Mean values and 68$\%$ credible intervals for  \lcdm\ $+A_L$ plus one of $N_{\textrm{eff}}$, $Y_{\textrm{He}}$, and $n_{\textrm{run}}$ for the MCMC chains fit to \planck\ TT. We use a prior of $\tau = 0.0506 \pm 0.0086$.  \label{tab:best-fit-lens+damp}}

\tablehead{
 \colhead{Parameter} & \colhead{$+ A_L$} & \colhead{$+ A_L + N_{\textrm{eff}}$} & \colhead{$+ A_L + n_{\textrm{run}}$} & \colhead{$+ A_L + Y_{\textrm{He}}$} 
 }
 
 \startdata
 $H_0$ & 69.11 $\pm$ 1.20 & 71.5 $\pm$ 3.8 & 69.14 $\pm$ 1.19 & 69.70 $\pm$ 1.49  \\
 $100*\omega_b$  & 2.265 $\pm$ 0.029 & 2.293 $\pm$ 0.052 & 2.271 $\pm$ 0.031 & 2.283 $\pm$ 0.040\\
 $\omega_c$ & 0.1164 $\pm$ 0.0025 & 0.1187 $\pm$ 0.0046 & 0.1164 $\pm$ 0.0025 & 0.1156 $\pm$ 0.0027 \\
 $10^9A_se^{-2\tau}$  & 1.8658 $\pm$ 0.0156 & 1.8770 $\pm$ 0.0157 & 1.8688 $\pm$ 0.0165 & 1.8696 $\pm$ 0.0170 \\
 $n_s$  & 0.9751 $\pm$ 0.0072 & 0.987 $\pm$ 0.020 & 0.9744 $\pm$ 0.0072 & 0.9822 $\pm$ 0.0122 \\
 $A_L$ & 1.259 $\pm$ 0.099 & 1.302 $\pm$ 0.117 & 1.270 $\pm$ 0.099 & 1.287 $\pm$ 0.107  \\
 $N_{\textrm{eff}}$ & --------- & 3.30 $\pm$ 0.40 & --------- & ---------  \\
 $n_{\textrm{run}}$ & --------- & --------- & -0.0050 $\pm$ 0.0076 & ---------  \\
 $Y_{\textrm{He}}$ & --------- & --------- & --------- & 0.261 $\pm$ 0.020  \\
 \hline
 $\Omega_mh^2$  & 0.1390 $\pm$ 0.0023 & 0.1415 $\pm$ 0.0048 & 0.1391 $\pm$ 0.0023 & 0.1384 $\pm$ 0.0024  \\
 $\Omega_m$  & 0.2915 $\pm$ 0.0148 & 0.279 $\pm$ 0.025 & 0.2915 $\pm$ 0.0147 & 0.2856 $\pm$ 0.0169  \\
 $\sigma_8$  & 0.7933 $\pm$ 0.0120 & 0.7992 $\pm$ 0.0157 & 0.7931 $\pm$ 0.0122 & 0.7936 $\pm$ 0.0124  \\
 \hline
 $\chi^2$  & 221.45& 220.91 & 221.14 & 220.99 \\
  $\chi^2_{\Lambda CDM + A_L} - \chi^2$  & 0 & 0.54 & 0.31 & 0.46 
 \enddata

 \vspace{-0.5cm}

\end{deluxetable*}

The results of adding $N_{\textrm{eff}}$, $n_{\textrm{run}}$, or $Y_{\textrm{He}}$ to \lcdm\ + $A_L$ when fitting to \planck\ TT data are shown in Table 5. From Table 5, it is clear that none of these result in a significant improvement to the fit. There is an increase of about 2.4 km$\textrm{s}^{-1}\textrm{Mpc}^{-1}$ in the preferred value of $H_0$ when allowing both $A_L$ and $N_{\textrm{eff}}$ to vary. This is accompanied by a roughly 300$\%$ increase in the uncertainty of $H_0$ placing the posterior for $H_0$ within 1$\sigma$ of the measured value by the cosmological distance ladder. However, when the \planck\ TE and EE power spectra are added to the fit, the constraint becomes $H_0 = 68.1 \pm 1.7$ km$\textrm{s}^{-1}\textrm{Mpc}^{-1}$, which corresponds to a 2.7$\sigma$ tension with the distance ladder preferred value for $H_0$. Therefore, \lcdm\ + $A_L$ + $N_{\textrm{eff}}$ is not a plausible resolution of the Hubble tension.

In this section, we have allowed various additional types of model freedom but found no substantial improvement over \lcdm\ + $A_L$; for whatever reason $A_L$ does seem to do a very effective job at relieving internal \planck\ tension. 

\section{Conclusions}
We test the impact of allowing phenomenological amplitudes for the Sachs-Wolfe, eISW, Doppler, and Polarization effects, which source the CMB temperature anisotropy, to vary when fitting to the \planck\ TT power spectrum. We find that allowing these amplitudes to vary results in only minimal improvement in the fit over standard \lcdm. Moreover, there are only minimal shifts in the preferred values of the \lcdm\ parameters when the amplitudes of these physical effects are varied. We conclude that \lcdm\ correctly accounts for each of these physical effects. 

Additionally, we test allowing multiple of these phenomenological amplitudes to vary simultaneously and find that allowing $A_{\textrm{SW}}$ and $A_{\textrm{Dop}}$ to vary together was the only combination that results in a significant improvement to the $\chi^2$ when fitting to \planck\ TT data. Howerver, we also show that allowing these two phenomenological amplitudes to vary simultaneously results in a significant degradation of the precision of $A_s$, which comes from the near rescaling of the power spectrum for multipoles $\ell > 400$ when $A_{\textrm{SW}}$ and $A_{\textrm{Dop}}$ are scaled in unison. When only multipoles $\ell > 800$ are included in the fit to the \planck\ TT spectrum, \lcdm\ + $A_{\textrm{SW}}$ + $A_{\textrm{Dop}}$ produces almost the same power spectrum as \lcdm\ + $A_L$. We conclude that \lcdm\ + $A_{\textrm{SW}}$ + $A_{\textrm{Dop}}$ is finding the \lcdm\ +  $A_L$ solution and therefore does not provide any new evidence for deviations from \lcdm\ predictions. 

From our tests where we vary $A_L$ and $A_{\textrm{eISW}}$ both simultaneously and separately, we quantitatively determine that the physical matter density is constrained primarily by the overall photon envelope with smaller contributions from both lensing and the eISW effect when fitting to \planck\ TT data. These findings are in line with \cite{Knox/etal:2020}. However, when only \planck\ TT $\ell > 800$ data is included, lensing provides the majority of the constraining power for the physical matter density. 

Finally, we varied both $A_L$ and one of $N_{\textrm{eff}}$, $n_{\textrm{run}}$, and $Y_{\textrm{He}}$ and fit to \planck\ TT data. All of these parameters impact the TT power spectrum at high $\ell$ meaning each of these parameter extensions provides a test of whether $A_L$ is fully able to resolve the internal tension between \planck\ TT $\ell \leq 800$ and \planck\ TT $\ell > 800$. We find no significant improvement in the fit over the \lcdm\ + $A_L$ case which suggests that there is little room for improvement from each of these effects. 

Allowing these phenomenological amplitudes for the physical effects that source the CMB temperature anisotropy to vary provides a new test of consistency of each of these physical effects with \lcdm\ predictions. While none of our new phenomenological tests provide evidence for deviations in the predictions made by \lcdm, this lack of deviations from \lcdm\ highlights that \lcdm\ is generally good at describing the very complex nature of the CMB temperature anisotropy with the caveat that there is a known \planck\ internal tension between $\ell \leq 800$ and $\ell > 800$. These tests suggest that any new model of cosmology will need to make similar predictions to \lcdm\ for the Sachs-Wolfe, Doppler, eISW and Polarization effects. 
\acknowledgments
This work was supported in part by NASA ROSES grants NNX17AF34G and 80NSSC19K0526. We acknowledge the use of the Legacy Archive for Microwave Background Data Analysis (LAMBDA), part of the High Energy Astrophysics Science Archive Center (HEASARC). HEASARC/LAMBDA is a service of the Astrophysics Science Division at the NASA Goddard Space Flight Center. This research project was conducted using computational resources at the Maryland Advanced Research Computing Center (MARCC). This work is based on observations obtained with Planck (http://www.esa.int/Planck), an ESA science mission with instruments and contributions directly funded by ESA Member States, NASA, and Canada. Figures 2, 3, and 4 were created using the Python package GetDist \citep{GetDist}. We thank Janet Weiland and Mario Aguilar Fa\'{u}ndez for helpful comments while completing the write up of this work.

\bibliography{cosmology,planck,act,spt}
\end{document}